# Stock Market Trend Analysis Using Hidden Markov Models

**Kavitha G**
School of Applied Sciences,
Hindustan University, Chennai, India.
E-mail: kavithateam@gmail.com

**\*Udhayakumar A**
School of Computing Sciences,
Hindustan University, Chennai, India
E-mail: aukumar71@gmail.com

**Nagarajan D**
Department of Information Technology, Math Section,
Salalah College of Technology,
Salalah, Sultanate of Oman
E-mail: drnagarajan75@gmail.com

*Abstract* — *Price movements of stock market are not totally random. In fact, what drives the financial market and what pattern financial time series follows have long been the interest that attracts economists, mathematicians and most recently computer scientists [17]. This paper gives an idea about the trend analysis of stock market behaviour using Hidden Markov Model (HMM). The trend once followed over a particular period will sure repeat in future. The one day difference in close value of stocks for a certain period is found and its corresponding steady state probability distribution values are determined. The pattern of the stock market behaviour is then decided based on these probability values for a particular time. The goal is to figure out the hidden state sequence given the observation sequence so that the trend can be analyzed using the steady state probability distribution($\pi$) values. Six optimal hidden state sequences are generated and compared. The one day difference in close value when considered is found to give the best optimum state sequence.*

*Keywords*-*Hidden Markov Model; Stock market trend; Transition Probability Matrix; Emission Probability Matrix; Steady State Probability distribution*

## I. INTRODUCTION

"A growing economy consists of prices falling, not rising", says Kel Kelly[9]. Stock prices change every day as a result of market forces. There is a change in share price because of supply and demand. According to the supply and demand, the stock price either moves up or undergoes a fall. Stock markets normally reflect the business cycle of the economy: when the economy grows, the stock market typically reflects this economic growth in an upward trend in prices. In contrast, when the economy slows, stock prices tend to be more mixed. Markets may take time to form bottoms or make tops, sometimes of two years or more. This makes it difficult to determine when the market hits a top or a bottom[3]. The Stock Market patterns are non-linear in nature, hence it is difficult to forecast future trends of the market behaviour.

In this paper, a method has been developed to forecast the future trends of the stock market. The Latent or hidden states, which determine the behaviour of the stock value, are usually invisible to the investor. These hidden states are derived from the emitted symbols. The emission probability depends on the current state of the HMM. Probability and Hidden Markov Model give a way of dealing with uncertainty. Many intelligent tasks are sequence finding tasks, with a limited availability of information. This naturally involves hidden states or strategies for dealing with uncertainty.

## II. LITERATURE SURVEY

In Recent years, a variety of forecasting methods have been proposed and implemented for the stock market analysis. A brief study on the literature survey is presented. Markov Process is a stochastic process where the probability at one time is only conditioned on a finite history, being in a certain state at a certain time. Markov chain is "Given the present, the future is independent of the past". HMM is a form of probabilistic finite state system where the actual states are not directly observable. They can only be estimated using observable symbols associated with the hidden states. At each time point, the HMM emits a symbol and changes a state with certain probability. HMM analyze and predict time series or time depending phenomena. There is not a one to one correspondence between the states and the observation symbols. Many states are mapped to one symbol and vice-versa.

Hidden Markov Model was first invented in speech recognition [12,13], but is widely applied to forecast stock market data. Other statistical tools are also available to make forecasts on past time series data. Box–Jenkins[2] used Time series analysis for forecasting and control. White[5,18,19] used Neural Networks for stock market forecasting of IBM daily stock returns. Following this, various studies reported on the effectiveness of alternative learning algorithms and prediction methods using ANN. To forecast the daily close and morning open price, Henry [6] used ARIMA model. But all these conventional methods had problems when non linearity exists in time series. Chiang et al.[4] have used ANN to forecast the end-of-year net asset value of mutual funds. Kim and Han [10] found that the complex dimensionality and buried noise of the stock market data makes it difficult to re-estimate the ANN parameters. Romahi and Shen [14] also found that ANN occasionally suffers from over fitting problem. They developed an evolving rule based expert system and obtained a method which is used to forecast financial market behaviour. There were also hybridization models effectively used to forecast financial behaviour. The drawback was requirement of expert knowledge. To overcome all these problems Hassan and Nath [15] used HMM for a better optimization. Hassan et al. [16] proposed a fusion model of HMM, ANN and GA for stock Market forecasting. In continuation of this, Hassan [7] combined HMM and fuzzy logic rules to improve the prediction accuracy on non-stationary stock data sets. Jyoti Badge[8] used technical indicators as an input variable instead of stock prices for analysis. Aditya Gupta and Bhuwan Dhingra[1] considered the fractional change in Stock value and the intra-day high and low values of the stock to train the continuous HMM**.** In the earlier studies, much research work had been carried out using various techniques and algorithms for training the model for forecasting or predicting the next day close value of the stock market, for which randomly generated Transition Probability Matrix (TPM), Emission Probability Matrix (EPM) and prior probability matrix have been considered.

\*Corresponding author

In this paper, the trend analysis of the stock market is found using Hidden Markov Model by considering the one day difference in close value for a particular period. For a given observation sequence, the hidden sequence of states and their corresponding probability values are found. The probability values of $\pi$ gives the trend percentage of the stock prices. Decision makers make decisions in case of uncertainty. The proposed approach gives a platform for decision makers to make decisions on the basis of the percentage of probability values obtained from the steady state probability distribution.

## III. RESEARCH SET UP

### A. Basics of HMM

HMM is a stochastic model where the system is assumed to be a Markov Process with hidden states. HMM gives better accuracy than other models. Using the given input values, the parameters of the HMM ($\lambda$) denoted by A, B and $\pi$ are found out.

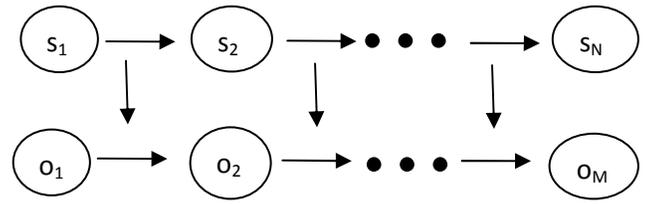

*Fig 1.* Trellis Diagram

HMM consists of

    A set of hidden or latent states (S)

    A set of possible output symbols (O)

    A state transition probability matrix (A)

➢ probability of making transition from one state to each of the other states

    Observation emission probability matrix (B)

➢ probability of emitting/observing a symbol at a particular state

    Prior probability matrix ($\pi$)

➢ probability of starting at a particular state

An HMM is defined as $\lambda=(S, O, A, B, \pi)$ where

    $S=\{s_1,s_2,\ldots,s_N\}$ is a set of N possible states

    $O=\{o_1,o_2,\ldots,o_M\}$ is a set of M possible observation symbols

    A is an NxN state Transition Probability Matrix (TPM)

    B is an NxM observation or Emission Probability Matrix (EPM)

    $\pi$ is an N dimensional initial state probability distribution vector

    and A, B and $\pi$ should satisfy the following conditions:

$$\sum_{j=1}^{N} a_{ij} = 1 \quad \text{where} \quad 1 \leq i \leq N;$$

$$\sum_{j=1}^{M} b_{ij} = 1 \quad \text{where} \quad 1 \leq i \leq N;$$

$$\sum_{i=1}^{N} \pi_i = 1 \quad \text{where} \quad \pi_i \geq 0$$

The main problems of HMM are: Evaluation, Decoding, and Learning.

*Evaluation problem*

Given the HMM $\lambda = \{A, B, \pi\}$ and the observation sequence $O = o_1 o_2 \ldots o_M$, the probability that model $\lambda$ has generated sequence $O$ is calculated.

Often this problem is solved by the Forward Backward Algorithm (Rabiner, 1989) (Rabiner, 1993).

*Decoding problem*

Given the HMM $\lambda = \{A, B, \pi\}$ and the observation sequence $O = o_1 o_2 \ldots o_M$, calculate the most likely sequence of hidden states that produced this observation sequence $O$.

Usually this problem is handled by Viterbi Algorithm (Rabiner, 1989) (Rabiner, 1993).

*Learning problem*

Given some training observation sequences $O = o_1 o_2 \ldots o_M$ and general structure of HMM (numbers of hidden and visible states), determine HMM parameters $\lambda = \{A, B, \pi\}$ that best fit training data.

The most common solution for this problem is Baum-Welch algorithm (Rabiner, 1989) (Rabiner, 1993) which is considered as the traditional method for training HMM.

In this paper, IBM daily close value data for a month period is considered.

Two observing symbols " I " and " D " have been used:

"I indicates Increasing" , " D indicates Decreasing ".

If Today's close value – Yesterday's close value > 0, then observing symbol is I

If Today's close value – Yesterday's close value < 0 then observing symbol is D

There are six hidden states assumed and are denoted by the symbol

S1, S2 , S3 , S4, S5, S6

where

    S1  indicates  "very low";
    S2  indicates  "low";
    S3  indicates  "moderate low"
    S4  indicates  "moderate high";
    S5  indicates  "high";
    S6  indicates  "very high".

The States are not directly observable. The situations of the stock market are considered hidden. Given a sequence of observation we can find the hidden state sequence that produced those observations.

### B. Database

The complete set of data for the proposed study has been taken from *yahoofinance.com*. The Table 1 given below shows the daily close value of the stock market:

*Table I. Daily close value for finding differences in one day, two days, three days, four days, five days, six days close value*

| S.NO | C.V | D.in.1 day CV | O.S | D.in.2 days CV | O.S | D.in.3 days CV | O.S | D.in.4 days CV | O.S | D.in.5 days CV | O.S | D.in.6 days CV | O.S |
|---|---|---|---|---|---|---|---|---|---|---|---|---|---|
| 1 | 77.91 | | | | | | | | | | | | |
| 2 | 77.39 | -0.52 | D | | | | | | | | | | |
| 3 | 76.5 | -0.89 | D | -1.41 | D | | | | | | | | |
| 4 | 75.86 | -0.64 | D | -1.53 | D | -2.05 | D | | | | | | |
| 5 | 77.45 | 1.59 | I | 0.95 | I | 0.06 | I | -0.46 | D | | | | |
| 6 | 79.33 | 1.88 | I | 3.47 | I | 2.83 | I | 1.94 | I | 1.42 | I | | |
| 7 | 79.51 | 0.18 | I | 2.06 | I | 3.65 | I | 3.01 | I | 2.12 | I | 1.6 | I |
| 8 | 79.15 | -0.36 | D | -0.18 | D | 1.7 | I | 3.29 | I | 2.65 | I | 1.76 | I |
| 9 | 79.95 | 0.8 | I | 0.44 | I | 0.62 | I | 2.5 | I | 4.09 | I | 3.45 | I |
| 10 | 78.56 | -1.39 | D | -0.59 | D | -0.95 | D | -0.77 | D | 1.11 | I | 2.7 | I |
| 11 | 79.07 | 0.51 | I | -0.88 | D | -0.08 | D | -0.44 | D | -0.26 | D | 1.62 | I |
| 12 | 77.4 | -1.67 | D | -1.16 | D | -2.55 | D | -1.75 | D | -2.11 | D | -1.93 | D |
| 13 | 77.28 | -0.12 | D | -1.79 | D | -1.28 | D | -2.67 | D | -1.87 | D | -2.23 | D |
| 14 | 77.95 | 0.67 | I | 0.55 | I | -1.12 | D | -0.61 | D | -2 | D | -1.2 | D |
| 15 | 77.33 | -0.62 | D | 0.05 | I | -0.07 | D | -1.74 | D | -1.23 | D | -2.62 | D |
| 16 | 76.7 | -0.63 | D | -1.25 | D | -0.58 | D | -0.7 | D | -2.37 | D | -1.86 | D |
| 17 | 77.73 | 1.03 | I | 0.4 | I | -0.22 | D | 0.45 | I | 0.33 | I | -1.34 | D |
| 18 | 77.07 | -0.66 | D | 0.37 | I | -0.26 | D | -0.88 | D | -0.21 | D | -0.33 | D |
| 19 | 77.9 | 0.83 | I | 0.17 | I | 1.2 | I | 0.57 | I | -0.05 | D | 0.62 | I |
| 20 | 75.7 | -2.2 | D | -1.37 | D | -2.03 | D | -1 | D | -1.63 | D | -2.25 | D |

C.V – Close value    ;  O.S – Observing symbol

D.in.1 day CV  - difference in 1 day close value;
D.in.2 days CV - difference in 2 days close value;
D.in.3 days CV - difference in 3 days close value;
D.in.4 days CV - difference in 4 days close value;
D.in.5 days CV - difference in 5 days close value;
D.in.6 days CV - difference in 6 days close value

## IV. CALCULATION

The various probability values of TPM, EPM and $\pi$ for difference in one day, two days, three days, four days, five days, six days close value close value are calculated as given below.

### A. Probability values of TPM, EPM and $\pi$ for difference in one day close value:

$$\begin{array}{c c}  & \begin{array}{c c c c c c} S1 & S2 & S3 & S4 & S5 & S6 \end{array} \\ \begin{array}{c} S1 \\ S2 \\ S3 \\ S4 \\ S5 \\ S6 \end{array} & \left[ \begin{array}{c c c c c c} 0 & 0 & 1 & 0 & 0 & 0 \\ 0 & 0 & 0.5 & 0.5 & 0 & 0 \\ 0 & 0.143 & 0.143 & 0 & 0.571 & 0.143 \\ 0.5 & 0 & 0.5 & 0 & 0 & 0 \\ 0.25 & 0.25 & 0.5 & 0 & 0 & 0 \\ 0 & 0 & 0 & 0.5 & 0 & 0.5 \end{array} \right] \end{array}$$

Fig 2. TPM

$$\begin{array}{c c} & \begin{array}{c c} I & D \end{array} \\ \begin{array}{c} S1 \\ S2 \\ S3 \\ S4 \\ S5 \\ S6 \end{array} & \left[ \begin{array}{c c} 0 & 1 \\ 0.5 & 0.5 \\ 0.71 & 0.29 \\ 0 & 1 \\ 0 & 1 \\ 1 & 0 \end{array} \right] \end{array}$$

Fig 3. EPM

Steady state probability distribution

$\pi = [0.06 \quad 0.11 \quad 0.39 \quad 0.11 \quad 0.22 \quad 0.11]$

Table II. Transition table with probability values for difference in one day close value

| TRANSITION OF STATES WITH OBSERVING SYMBOLS | S1 | | S2 | | S3 | | S4 | | S5 | | S6 | |
|---|---|---|---|---|---|---|---|---|---|---|---|---|
| | I | D | I | D | I | D | I | D | I | D | I | D |
| S1 | 0 | 0 | 0 | 0 | 0 | 1 | 0 | 0 | 0 | 0 | 0 | 0 |
| S2 | 0 | 0 | 0 | 0 | 0 | 0.5 | 0.5 | 0 | 0 | 0 | 0 | 0 |
| S3 | 0 | 0 | 0 | 0.1429 | 0 | 0.1429 | 0 | 0 | 0.5714 | 0 | 0.1429 | 0 |
| S4 | 0 | 0.5 | 0 | 0 | 0 | 0.5 | 0 | 0 | 0 | 0 | 0 | 0 |
| S5 | 0 | 0.25 | 0 | 0.25 | 0 | 0.5 | 0 | 0 | 0 | 0 | 0 | 0 |
| S6 | 0 | 0 | 0 | 0 | 0 | 0 | 0 | 0 | 0 | 0 | 1 | 0 |

### B. Probability values of TPM, EPM and $\pi$ for difference in two days close value:

$$\begin{array}{c c} & \begin{array}{c c c c c c} S1 & S2 & S3 & S4 & S5 & S6 \end{array} \\ \begin{array}{c} S1 \\ S2 \\ S3 \\ S4 \\ S5 \\ S6 \end{array} & \left[ \begin{array}{c c c c c c} 0.4 & 0 & 0.4 & 0.2 & 0 & 0 \\ 0.33 & 0.33 & 0.33 & 0 & 0 & 0 \\ 0.33 & 0.17 & 0.5 & 0 & 0 & 0 \\ 0 & 0 & 0 & 0 & 0 & 1 \\ 0 & 1 & 0 & 0 & 0 & 0 \\ 0 & 0 & 0 & 0 & 1 & 0 \end{array} \right] \end{array}$$

Fig 4. TPM

$$\begin{array}{c c} & \begin{array}{c c} I & D \end{array} \\ \begin{array}{c} S1 \\ S2 \\ S3 \\ S4 \\ S5 \\ S6 \end{array} & \left[ \begin{array}{c c} 0.6 & 0.4 \\ 0.33 & 0.67 \\ 0.5 & 0.5 \\ 1 & 0 \\ 0 & 1 \\ 0 & 1 \end{array} \right] \end{array}$$

Fig 5. EPM

Steady state probability distribution

$\pi = [0.29 \quad 0.18 \quad 0.35 \quad 0.06 \quad 0.06 \quad 0.06]$

*Table III. Transition table with probability values for difference in two days close values*

| TRANSITION OF STATES WITH OBSERVING SYMBOLS | S1 | | S2 | | S3 | | S4 | | S5 | | S6 | |
|---|---|---|---|---|---|---|---|---|---|---|---|---|
| | I | D | I | D | I | D | I | D | I | D | I | D |
| S1 | 0 | 0.4 | 0 | 0 | 0.4 | 0 | 0.2 | 0 | 0 | 0 | 0 | 0 |
| S2 | 0 | 0.33 | 0 | 0.33 | 0.33 | 0 | 0 | 0 | 0 | 0 | 0 | 0 |
| S3 | 0 | 0.33 | 0 | 0.167 | 0.5 | 0 | 0 | 0 | 0 | 0 | 0 | 0 |
| S4 | 0 | 0 | 0 | 0 | 0 | 0 | 0 | 0 | 0 | 0 | 1 | 0 |
| S5 | 0 | 0 | 0 | 1 | 0 | 0 | 0 | 0 | 0 | 0 | 0 | 0 |
| S6 | 0 | 0 | 0 | 0 | 0 | 0 | 0 | 0 | 0 | 1 | 0 | 0 |

*Table IV. Transition table with probability values for difference in three days close values*

| TRANSITION OF STATES WITH OBSERVING SYMBOLS | S1 | | S2 | | S3 | | S4 | | S5 | | S6 | |
|---|---|---|---|---|---|---|---|---|---|---|---|---|
| | I | D | I | D | I | D | I | D | I | D | I | D |
| S1 | 0 | 0 | 0 | 0.5 | 0.5 | 0 | 0 | 0 | 0 | 0 | 0 | 0 |
| S2 | 0 | 0 | 0 | 0.25 | 0 | 0.75 | 0 | 0 | 0 | 0 | 0 | 0 |
| S3 | 0 | 0.2 | 0 | 0.2 | 0 | 0.2 | 0.2 | 0 | 0 | 0 | 0.2 | 0 |
| S4 | 0 | 0.5 | 0 | 0.5 | 0 | 0 | 0 | 0 | 0 | 0 | 0 | 0 |
| S5 | 0 | 0 | 0 | 0 | 0 | 0 | 1 | 0 | 0 | 0 | 0 | 0 |
| S6 | 0 | 0 | 0 | 0 | 0 | 0 | 0 | 0 | 0.5 | 0 | 0.5 | 0 |

**C. Probability values of TPM, EPM and π for difference in three days close value:**

$$\begin{array}{c}\phantom{S1}\quad S1\quad S2\quad S3\quad S4\quad S5\quad S6\\ \begin{array}{c}S1\\ S2\\ S3\\ S4\\ S5\\ S6\end{array}\begin{bmatrix}0 & 0.5 & 0.5 & 0 & 0 & 0\\ 0 & 0.25 & 0.75 & 0 & 0 & 0\\ 0.2 & 0.2 & 0.2 & 0.2 & 0 & 0.2\\ 0.5 & 0.5 & 0 & 0 & 0 & 0\\ 0 & 0 & 0 & 1 & 0 & 0\\ 0 & 0 & 0 & 0 & 0.5 & 0.5\end{bmatrix}\end{array}$$

*Fig 6. TPM*

$$\begin{array}{c}\phantom{S1}\quad I\quad\ \ D\\ \begin{array}{c}S1\\ S2\\ S3\\ S4\\ S5\\ S6\end{array}\begin{bmatrix}0.5 & 0.5\\ 0 & 1\\ 0.4 & 0.6\\ 0 & 1\\ 1 & 0\\ 1 & 0\end{bmatrix}\end{array}$$

*Fig 7. EPM*

Steady state probability distribution
$\pi = [0.13\ \ 0.25\ \ 0.31\ \ 0.13\ \ 0.06\ \ 0.13]$

**D. Probability values of TPM, EPM and π for difference in four days close value:**

$$\begin{array}{c}\phantom{S1}\quad S1\quad\ \ S2\quad\ \ S3\quad\ \ S4\quad\ \ S5\quad\ S6\\ \begin{array}{c}S1\\ S2\\ S3\\ S4\\ S5\\ S6\end{array}\begin{bmatrix}0.33 & 0.33 & 0.33 & 0 & 0 & 0\\ 0 & 0 & 0.33 & 0.67 & 0 & 0\\ 0.67 & 0 & 0 & 0 & 0.33 & 0\\ 0 & 1 & 0 & 0 & 0 & 0\\ 0 & 0 & 0 & 0 & 0 & 1\\ 0 & 0.33 & 0 & 0 & 0 & 0.67\end{bmatrix}\end{array}$$

*Fig 8. TPM*

$$\begin{array}{c}\phantom{S1}\quad I\quad\ \ D\\ \begin{array}{c}S1\\ S2\\ S3\\ S4\\ S5\\ S6\end{array}\begin{bmatrix}0 & 1\\ 0.67 & 0.33\\ 0.33 & 0.67\\ 0 & 1\\ 1 & 0\\ 0.67 & 0.33\end{bmatrix}\end{array}$$

*Fig 9. EPM*

Steady state probability distribution
$$\pi = [0.04 \quad 0.04 \quad 0.04 \quad 0.13 \quad 0.07 \quad 0.04]$$

Table V.  Transition table with probability values for difference in four days close values

| TRANSITION OF STATES WITH OBSERVING SYMBOLS | S1 | | S2 | | S3 | | S4 | | S5 | | S6 | |
|---|---|---|---|---|---|---|---|---|---|---|---|---|
| | I | D | I | D | I | D | I | D | I | D | I | D |
| S1 | 0 | 0.33 | 0 | 0.33 | 0 | 0.33 | 0 | 0 | 0 | 0 | 0 | 0 |
| S2 | 0 | 0 | 0 | 0 | 0 | 0.33 | 0.67 | 0 | 0 | 0 | 0 | 0 |
| S3 | 0 | 0.67 | 0 | 0 | 0 | 0 | 0 | 0 | 0.33 | 0 | 0 | 0 |
| S4 | 0 | 0 | 0 | 1 | 0 | 0 | 0 | 0 | 0 | 0 | 0 | 0 |
| S5 | 0 | 0 | 0 | 0 | 0 | 0 | 0 | 0 | 0 | 0 | 1 | 0 |
| S6 | 0 | 0 | 0 | 0.33 | 0 | 0 | 0 | 0 | 0 | 0 | 0.67 | 0 |

E.  Probability values of TPM, EPM and π for difference in five days close value:

$$\begin{array}{c}\quad\;\; S1 \quad S2 \quad S3 \quad S4 \quad S5 \quad S6 \\ \begin{array}{c}S1\\S2\\S3\\S4\\S5\\S6\end{array}\begin{bmatrix} 0.5 & 0.25 & 0.25 & 0 & 0 & 0 \\ 1 & 0 & 0 & 0 & 0 & 0 \\ 0.33 & 0 & 0.67 & 0 & 0 & 0 \\ 0 & 0.5 & 0 & 0 & 0.5 & 0 \\ 0 & 0 & 0 & 0 & 0.5 & 0.5 \\ 0 & 0 & 0 & 1 & 0 & 0 \end{bmatrix}\end{array}$$

Fig 10.  TPM

$$\begin{array}{c}\quad\;\; I \quad\;\; D \\ \begin{array}{c}S1\\S2\\S3\\S4\\S5\\S6\end{array}\begin{bmatrix} 0.25 & 0.75 \\ 0 & 1 \\ 0 & 1 \\ 0.5 & 0.5 \\ 1 & 0 \\ 1 & 0 \end{bmatrix}\end{array}$$

Fig 11.  EPM

Steady state probability distribution
$$\pi = [0.29 \quad 0.14 \quad 0.21 \quad 0.14 \quad 0.14 \quad 0.07]$$

Table VI.  Transition table with probability values for difference in five days close values

| TRANSITION OF STATES WITH OBSERVING SYMBOLS | S1 | | S2 | | S3 | | S4 | | S5 | | S6 | |
|---|---|---|---|---|---|---|---|---|---|---|---|---|
| | I | D | I | D | I | D | I | D | I | D | I | D |
| S1 | 0.5 | 0 | 0 | 0.25 | 0.25 | 0 | 0 | 0 | 0 | 0 | 0 | 0 |
| S2 | 0 | 1 | 0 | 0 | 0 | 0 | 0 | 0 | 0 | 0 | 0 | 0 |
| S3 | 0 | 0.33 | 0 | 0 | 0 | 0.67 | 0 | 0 | 0 | 0 | 0 | 0 |
| S4 | 0 | 0 | 0 | 0.5 | 0 | 0 | 0 | 0 | 0.5 | 0 | 0 | 0 |
| S5 | 0 | 0 | 0 | 0 | 0 | 0 | 0 | 0 | 0.5 | 0 | 0.5 | 0 |
| S6 | 0 | 0 | 0 | 0 | 0 | 0 | 1 | 0 | 0 | 0 | 0 | 0 |

F.  Probability values of TPM, EPM and π for difference in six days close value:

$$\begin{array}{c}\quad\;\; S1 \quad S2 \quad S3 \quad S4 \quad S5 \quad S6 \\ \begin{array}{c}S1\\S2\\S3\\S4\\S5\\S6\end{array}\begin{bmatrix} 0.5 & 0.5 & 0 & 0 & 0 & 0 \\ 0.5 & 0 & 0.5 & 0 & 0 & 0 \\ 0 & 0 & 0 & 1 & 0 & 0 \\ 1 & 0 & 0 & 0 & 0 & 0 \\ 0.33 & 0 & 0 & 0 & 0.33 & 0.33 \\ 0 & 0 & 0 & 0 & 0.5 & 0.5 \end{bmatrix}\end{array}$$

Fig 12.  TPM

$$\begin{array}{c}\quad\;\; I \quad\;\; D \\ \begin{array}{c}S1\\S2\\S3\\S4\\S5\\S6\end{array}\begin{bmatrix} 0 & 1 \\ 0 & 1 \\ 1 & 0 \\ 0 & 1 \\ 0.67 & 0.33 \\ 1 & 0 \end{bmatrix}\end{array}$$

Fig 13.  EPM

Steady state probability distribution
$$\pi = [0.31 \quad 0.15 \quad 0.08 \quad 0.08 \quad 0.23 \quad 0.15]$$

*Table VII.* **Transition table with probability values for difference in six days close values**

| TRANSITION OF STATES WITH OBSERVING SYMBOLS | S1 | | S2 | | S3 | | S4 | | S5 | | S6 | |
|---|---|---|---|---|---|---|---|---|---|---|---|---|
| | I | D | I | D | I | D | I | D | I | D | I | D |
| S1 | 0 | 0.5 | 0 | 0.5 | 0 | 0 | 0 | 0 | 0 | 0 | 0 | 0 |
| S2 | 0 | 0.5 | 0 | 0 | 0 | 0.5 | 0 | 0 | 0 | 0 | 0 | 0 |
| S3 | 0 | 0 | 0 | 0 | 0 | 0 | 1 | 0 | 0 | 0 | 0 | 0 |
| S4 | 0 | 1 | 0 | 0 | 0 | 0 | 0 | 0 | 0 | 0 | 0 | 0 |
| S5 | 0 | 0.33 | 0 | 0 | 0 | 0 | 0 | 0 | 0.33 | 0 | 0.33 | 0 |
| S6 | 0 | 0 | 0 | 0 | 0 | 0 | 0 | 0 | 0.5 | 0 | 0.5 | 0 |

The MATLAB function "Hmmgenerate" is used to generate a random sequence of emission symbols and states. The length of both sequence and states to be generated is denoted by L.

The HMM matlab toolbox syntax is :

[Sequence, States] = Hmmgenerate ( L , TPM, EPM) , see [11]

For instance,

If the Input is given as,

TPM = [0 0 1 0 0 0; 0 0 0.5 0.5 0 0; 0 0.143 0.143 0 0.571 0.143; 0.5 0 0.5 0 0 0; 0.25 0.25 0.5 0 0 0; 0 0 0 0.5 0 0.5];

EPM = [0 1;0.5 0.5; 0.71 0.29;0 1;0 1;1 0];

[sequence,states] = hmmgenerate(7, TPM, EPM)

```
'Sequence Symbols',{'I','D'},...
    'Statenames',{'very
low';'low';'moderate low';'moderate
high';'high';'very high'}
```

Then the Output of few randomly generated sequences and states is given below:

Sequence:  $\varepsilon \to I \to D \to D \to I \to I \to I \to I$

states   :     S3    S2    S3    S6    S6    S6    S6

sequence :  $\varepsilon \to D \to I \to D \to D \to I \to I \to I$

states   :     S3    S3    S5    S1    S3    S2    S3

where '$\varepsilon$' denotes the start symbol.

The fitness function used for finding the fitness value of sequence of states is defined by

$$\text{Fitness} = \frac{1}{\sum compare(i,j)}$$

## V. DISCUSSION

Using the Iterative procedure, for each TPM and EPM framed we get an optimum sequence of states generated. The length of the sequence generated is taken as L=7, for instance.

The optimum sequence of states obtained from the one day difference TPM and EPM is

1. $\varepsilon \to D \to I \to D \to I \to D \to I$
       S1    S3    S5    S3    S5    S3    S5

Similarly, we get 5 more such optimum sequences of states for 2 day difference, 3 day difference, 4 day difference, 5 day difference, 6 day difference TPM and EPM respectively as follows:

2. $\varepsilon \to I \to D \to D \to I \to D \to D$
       S1    S3    S1    S1    S3    S1    S1

3. $\varepsilon \to D \to D \to I \to D \to I \to I$
       S1    S2    S3    S4    S1    S3    S4

4. $\varepsilon \to D \to I \to D \to I \to D \to D$
       S1    S2    S4    S2    S4    S2    S3

5. $\varepsilon \to D \to D \to I \to I \to D \to D$
       S1    S2    S1    S1    S1    S2    S1

6. $\varepsilon \to D \to D \to I \to D \to D \to D$
       S1    S2    S3    S4    S1    S2    S3

Using the fitness function we compute the fitness value for each of the optimum sequence of states obtained.

*Table VIII. Comparison of Six Optimum State Sequences*

| S.No. | Comparison of 6 optimum sequence of states | Calculated value | Fitness = $\frac{1}{\sum compare(i,j)}$ |
|---|---|---|---|
| 1. | (1,2) + (1,3) + (1,4) + (1,5) + (1,6) | 1 | 1 |
| 2. | (2,1) + (2,3) + (2,4) + (2,5) + (2,6) | 1.29 | 0.76 |
| 3. | (3,1) + (3,2) + (3,4) + (3,5) + (3,6) | 1.86 | 0.54 |
| 4. | (4,1) + (4,2) + (4,3) + (4,5) + (4,6) | 1.43 | 0.70 |
| 5. | (5,1) + (5,2) + (5,3) + (5,4) + (5,6) | 2.14 | 0.47 |
| 6. | (6,1) + (6,2) + (6,3) + (6,4) + (6,5) | 2.14 | 0.47 |

## VI. CONCLUSION

In this paper, results are presented using Hidden Markov Model to find the trend of the stock market behavior. The highest is the fitness value, the better is the performance of the particular sequence. One day difference in close value when considered is found to give the best optimum sequence. It is observed that at any point of time over years, if the stock market behaviour pattern is the same then we can observe the same steady state probability values as obtained in one day difference of close value, which clearly determines the behavioural pattern of the stock market.